\documentclass[aps,amssymb,amsmath,twocolumn,secnumarabic]{revtex4}

\usepackage[dvips]{graphicx}
\usepackage{amsfonts}
\usepackage{amsmath}
\usepackage{mathrsfs}
\usepackage{color}
\usepackage[arrow, matrix, curve]{xy}
\xymatrixcolsep{0.5cm} \xymatrixrowsep{0.5cm}
\newcommand{\beqa}{\begin{eqnarray}}
\newcommand{\eeqa}{\end{eqnarray}}
\newcommand{\beq}{\begin{equation}}
\newcommand{\eeq}{\end{equation}}
\renewcommand{\l}{\lambda}

\newcommand{\D}{\Delta}
\newcommand{\tFo}[3]{\ _2F_1\left( \begin{array}{c} #1 \\  #2
\end{array}  ; {#3}\right)}

\begin{document}

\title{Superintegrability of $d$-dimensional Conformal Blocks}

\author{Mikhail Isachenkov$^{a}$ and Volker Schomerus$^{b}$}

\affiliation{$^a$Department of Particle Physics and Astrophysics, Weizmann Institute of Science, Rehovot 7610001, Israel\\
$^b$DESY Theory Group, DESY Hamburg, Notkestrasse 85, D-22603 Hamburg,
Germany}

\date{Feb 2016}

\begin{abstract}
We observe that conformal blocks of scalar 4-point functions in
a $d$-dimensional conformal field theory can mapped to eigenfunctions 
of a 2-particle hyperbolic Calogero-Sutherland Hamiltonian. The latter 
describes two coupled P\"oschl-Teller particles. Their interaction, whose
strength depends smoothly on the dimension $d$, is known to be superintegrable. 
Our observation enables us to exploit the rich mathematical literature  
on Calogero-Sutherland models in deriving various results for conformal 
field theory. These include an explicit construction of conformal blocks 
in terms of Heckman-Opdam hypergeometric functions and a remarkable 
duality that relates the blocks of theories in different dimensions. 
\\
\vskip 0.1cm \hskip -0.3cm
\end{abstract}

\maketitle

\vspace*{-9.9cm} \noindent
{\tt DESY 16-nnn\\
\tt WIS/02/16-FEB-DPPA\\

   \hspace*{12.7cm} \phantom{arXiveyymm.nnnn}}\\[6.7cm]

\section{Introduction}

Conformal quantum field theories (CFTs) play an important role for
modern theoretical physics. In statistical physics they describe the
universal  behavior of second order phase transitions. At the same time,
CFTs also provide a window into interacting and strongly coupled quantum field
theories which are very difficult to access otherwise. In $d=2$
dimensions, the global conformal algebra is extended to an infinite-dimensional 
symmetry. This was exploited to construct
many such models, paving the way for numerous applications
in diverse areas of physics and mathematics.

While the symmetry enhancement of 2-dimensional CFT is certainly
helpful, it may not be decisive. In fact, CFTs in any dimension
$d$ are very strongly constrained by global conformal symmetry.
Within the so-called conformal bootstrap programme, the solution
of CFTs can be reduced to certain integral equations, the
{\it crossing symmetry} constraints \cite{Ferrara:1973yt,
Polyakov:1974gs,Mack:1975jr}. These provide a system of equations
for the dynamical coefficients in the operator product expansion
involving only the kinematically determined crossing kernel, i.e.\
group theoretic data. And indeed, recent numerical studies of the
crossing symmetry equations, in particular for the conformal
Ising model in $d=3$ dimensions, have provided ample new
precision data on this model, see \cite{ElShowk:2012ht} and
references therein.

Analytical progress is lagging behind partly because it is
restricted to certain limits in which there exists sufficient
control of the kinematical input. This is what our work
addresses. We will focus on the group theoretic building
blocks of scalar 4-point functions, the so-called {\it
conformal blocks} that underly the entire bootstrap
programme. Partial waves can be characterized through a
second order differential equation \cite{Dolan:2003hv}.
So far, a construction of solutions of conformal Casimir
equations in terms of hypergeometric functions is only
known in even integer dimensions, where they can be
obtained from Gauss hypergeometric functions.

Our main observation is that the Casimir equation for
conformal blocks in $d$ dimensions may be transformed
into the eigenvalue problem for a Calogero-Sutherland (CS) Hamiltonian,
whose eigenfunctions are given by Heckman-Opdam (HO) hypergeometric
functions \cite{Heckman1987}. Thereby we connect the poorly developed
theory of conformal blocks to integrability and the modern theory
of special functions. The relevant CS Hamiltonian
turns out to be  superintegrable, i.e. it possesses an additional Runge-Lenz-like
integral of motion \cite{Kuznetsov:1995xn}. The latter is part of the (degenerate) 
double affine Hecke algebra (DAHA) \cite{Cherednik2005} which provides an extremely 
powerful algebraic underpinning, introduces a distinguished q-deformation
and bridges to the dual Ruijsenaars-Schneider (RS) model
\cite{Ruijsenaars:1986vq}. This leads to a wealth of interesting
relations for conformal blocks some of which we shall
touch upon below.

The plan of this paper is as follows. In the next section we
will briefly review the characterization of conformal blocks through
the conformal Casimir equation. For pedagogical
reasons we shall then explore our general theme in $d=2$ where
the relevant CS model decouples into two P\"oschl-Teller systems.
These are known to be solvable through hypergeometric functions.
Then we turn to the $d$-dimensional problem in section 4 and
explain how the two P\"oschl-Teller systems are coupled in order
to describe conformal blocks in $d$-dimensional conformal
field theory. The known eigenfunctions of the resulting CS
Hamiltonian are used in section 5 to construct conformal blocks
from a q-deformed version of HO hypergeometric functions. We
conclude by highlighting a few applications of known mathematical
results, many of them quite recent, to the conformal bootstrap
programme.

\section{Conformal Partial Waves}

In this section want to set up the problem by briefly reviewing some
material from \cite{Dolan:2003hv}. The correlation function of four
scalar conformal primary fields of weight $\Delta_i, i = 1,\dots,4$
in a $d$-dimensional conformal field theory can be decomposed as
\beqa
& & \hspace*{-1cm} \nonumber
\langle\phi_1(x_1)\phi_2(x_2)\phi_3(x_3)\phi_4(x_4)\rangle \\
& & \hspace*{-8mm}
= \frac{1}{x_{12}^{\frac12(\D_1+\D_2)}x_{34}^{\frac12(\D_3+\D_4)}}
\left(\frac{x_{14}}{x_{24}}\right)^a
\left(\frac{x_{14}}{x_{13}}\right)^b  G(z,\bar z)
\eeqa
with $x_{ij} = x_i - x_j$ and $2a=\D_2-\D_1, 2b= \D_3-\D_4$.
The conformal invariants $z, \bar z$ were introduced to
parameterize the more familiar cross ratios as
\beqa
\frac{x^2_{12}x^2_{34}}{x^2_{13}x^2_{24}} & = & z\bar z   \\
\frac{x^2_{14}x^2_{23}}{x^2_{13}x^2_{24}} & = & (1-z)(1-\bar z)  \ .
\eeqa
For a Euclidean theory, $z,\bar z$ are complex variables. 
The function $G$ receives contributions from all the primary fields
that can appear in the operator product expansion of the field $\phi_1$
and $\phi_2$
\beq
G(z,\bar z) = \sum_{\D,l} \lambda^{12}_l(\D) \lambda^{34}_l(\Delta)
G_{\D,l}(z,\bar z)\ .
\eeq
This expansion separates the dynamically determined coefficients $\lambda$
of the operator product from the kinematic {\em conformal blocks}
$G_{\Delta,l}$. The latter are eigenfunctions of the conformal Laplacian
$D^2_\epsilon$,
\beq \label{CPWdeq}
 D^2_\epsilon G(z,\bar z) = \frac12 C_{\D,l} G(z,\bar z)
\eeq
with eigenvalues
\beq \label{CPWev}
 C_{\D,l} = \Delta(\Delta-d) +l (l+d-2) \ 
\eeq
and subject to an additional boundary condition that selects a 
unique (up to normalization) combination of solutions. The form 
of the conformal Laplacian can be worked out easily, see
e.g.\ \cite{Dolan:2003hv},
\beq
 D^2_\epsilon := D^2 + \overline{D}^2 +
 \epsilon\left[ \frac{z \overline{z}}{\overline{z}-z} \left( \overline{\partial} -
 \partial\right)
 \label{Diffop}  +(z^{2}\partial-\overline{z}^{2}\overline{\partial})\right]
\eeq
where $\epsilon = d-2$ and
\beq
D^2 = z^{2}(1-z)\partial^{2} -(a+b+1)z^{2}\partial\ - ab z.
\eeq
$\overline{D}^2$ is defined similarly in terms of $\bar z$. In $d=2$
dimensions the Hamiltonian splits into sum of two independent pieces and
the corresponding eigenvalue equations are straightforwardly related to
hypergeometric differential equations. Our main goal in this work is to
solve the eigenvalue problem for the conformal Laplacian in terms of some
known special functions.

\section{P\"oschl-Teller potential}

In order to get a bit more insight into the structure of the eigenvalue
problem for the conformal Laplacian we will temporarily set $d=2$. The
Laplacian then decomposes into a sum of operators acting on $z$ and
$\bar z$ only and we shall focus on the eigenvalue problem for $D^2$.
This problem leads to the following second order differential equation
$$
D^2 G(z) = h(h-1)G(z)
$$
Now let us now define a new function which is related to $G$ by a
`gauge transformation' of the form
\beq \label{PTGT}
\psi(x) := \frac{(z-1)^{\frac{a+b}{2}+\frac{1}{4}}}{\sqrt{z}}
\ G(z)
\eeq
where the coordinates $z$ and $x$ are related by
\beq \label{zxrel}
z= - \sinh^{-2}\frac{x}{2}\ .
\eeq
Note that this relation maps the complex $z$-plane to a semi-infinite
strip with {\it Re}$\,x \geq 0$ and {\it Im}$\,x \in [0,\pi]$. Inserting these
relations it is easy to see that the function $\psi$ is an eigenfunction of
the P\"oschl-Teller Hamiltonian with potential
\beq \label{PTpot}
 V^{(a,b)}_{\text{PT}}(x) = \frac{(a+b)^2-\frac{1}{4}}{\sinh^2x} -
   \frac{ab}{\sinh^2(x/2)}
\eeq
for the eigenvalue $\varepsilon := 2mE/\hbar^2 = - (2h-1)^2/4$. The original Schr\"odinger problem 
studied by P\"oschl and Teller in \cite{Poschl:1933zz} was a trigonometric version of 
eq.\ \eqref{PTpot}. After such rotation to $y=ix$ the associated Schr\"odinger problem
describes a particle that is confined to the interval $y \in [0,\pi]$.
The Hamilton operator possesses a discrete spectrum with eigenfunctions
given by ordinary Jacobi polynomials.

The hyperbolic version we are dealing with here is also referred to as
P\"oschl-Teller Hamiltonian of second kind. It describes a particle on
the half-line $x\geq 0$. Since the potential falls off to zero for large
$x$, the Hamiltonian has a continuous part in its spectrum. The eigenfunctions
are given by
$$\psi_h(x) \sim z^{h-\frac12} (z-1)^{\frac{a+b}{2}+\frac{1}{4}}
   \tFo{h+a,h+b}{2h}{z}\ . $$
Before we move on we stress that the P\"oschl-Teller problem is
related to some classical theory of special functions. Let us describe
this for the trigonometric case in which eigenfunctions are classical
Jacobi polynomials. Like all other hypergeometric orthogonal polynomials
in a single variable, Jacobi polynomials are obtained from a degeneration
of the so-called Askey-Wilson polynomials. The latter may be constructed
from the q-deformed version $_4\Phi_3$ of the hypergeometric function
$_4F_3$ by specializing its parameters, see e.g.\ \cite{Koornwinder:2012}.
Of course, all these relations can be lifted to the hyperbolic theory,
i.e.\ from polynomials to functions.

\section{Calogero-Sutherland potential}

Historically, the Schr\"odinger problem for the P\"oschl-Teller potential
was solved through the relation with the hypergeometric differential
equation. But today it is much more interesting to look at the relation
in the opposite direction. Following work of Calogero, Moser and
Sutherland in the early 1970s, see \cite{Calogero:1970nt,Moser:1975qp,
Sutherland:1971ks}, the solvable P\"oschl-Teller problem has been
generalized in several directions. In particular it was understood that
the P\"oschl-Teller potential is just the simplest example of a large
family of superintegrable Schr\"odinger problems involving multiple
particles. The relevant potentials are associated with reflection groups
and give rise to so-called (trigonometric or hyperbolic)
Calogero-(Moser)-Sutherland models \cite{Olshanetsky:1983wh}.

In the last section we recalled that the Casimir equation for blocks in
2-dimensional chiral conformal field theory is equivalent to
the P\"oschl-Teller problem. Our main claim is that this extends to the
full Casimir equation for conformal blocks in $d$ dimensions.
In complete analogy to the discussion above, it turns out that the
Casimir equation is equivalent to the hyperbolic CS
model for reflection group {\it BC}$_2$. Its potential is given by
\beqa V^{(a,b,\epsilon)}_{\text{CS}}(x_1,x_2) & = &
V^{(a,b)}_{\text{PT}}(x_1) +
V^{(a,b)}_{\text{PT}}(x_2) + \nonumber  \\[2mm]
& & \hspace*{-7mm} + \frac{\epsilon(\epsilon-2)}
{8\sinh^2\frac{x_1-x_2}{2}} + \frac{\epsilon(\epsilon-2)}
{8\sinh^2 \frac{x_1+x_2}{2}} \ . \label{CSpot}
\eeqa
It is built from two P\"oschl-Teller systems with an interaction term
whose coupling explicitly depends on the dimension $d$. The six terms of this
potential reflect the six positive roots of the {\it BC}$_2$
root system. To relate the associated Schr\"odinger problem on the 
{\it BC}$_2$ Weyl chamber with the eigenvalue equation \eqref{CPWdeq} for 
the conformal Laplacian we generalize the gauge transformation \eqref{PTGT} 
to become
\beq \label{CSGT}
\psi(x_1,x_2) := \prod_i
\frac{(z_i-1)^{\frac{a+b}{2}+\frac{1}{4}}}{z_i^{\frac12+\frac{\epsilon}{2}}}
|z_1-z_2|^{\frac{\epsilon}{2}} G(z_1,z_2)
\eeq
where $z_1 = z$ and $z_2 = \bar z$. It is not difficult to verify that
this gauge transformation, along with the relation
\beq \label{zixirel}
z_i= - \sinh^{-2}\frac{x_i}{2}\ .
\eeq
between the coordinates $z_i$ and $x_i$, turns the conformal Laplacian
into the CS Hamiltonian for the potential \eqref{CSpot}, with the eigenvalue $\varepsilon = - d(d-2)/4-(C_{\D,l}+1)/2$. The appearance
of the {\it BC}$_2$ root lattice possesses a natural explanation in the
corresponding harmonic analysis formulation of the problem
\cite{Oblomkov2004153}, where it enters as a projection of the root
lattice of the conformal algebra $\mathfrak{so}(1,d+1),d\geq 5,$ to
the 2-dimensional plane spanned by the Cartan generators of an
embedded $\mathfrak{so}(1,3)$.

Just as the 1-dimensional P\"oschl-Teller problem is exactly solvable, so are
the higher-dimensional CS extensions and hence, by the relation
\eqref{CSGT}, the eigenvalue problem for the conformal Laplacian. Let us note that
the coupling constant in front of the interaction term is $\varepsilon= d-2$. In
$d=2$ dimensions, we are just dealing with two independent integrable P\"oschl-Teller
systems. Going away from $d=2$ introduces a new coupling in the potential. It is quite
remarkable that this coupling is also integrable. One may notice that in $d=4$ dimension
the interaction terms vanishes once again. This implies that 2d and 4d conformal blocks
are simply related by a gauge transformation. The latter switches between
bosonic/fermionic statistics of the wave function. For conformal blocks, the
simple relation between $d=2$ and $d=4$ dimensions is indeed consistent with the
standard expressions \cite{Dolan:2003hv}.

\section{Some applications}

What makes the observed relation between conformal blocks and
the CS Hamiltonian interesting are the connections of
the latter with integrability and the modern theory of special functions.
The integrability of the CS model can be established
using so-called {\em Dunkl operators}, i.e.\ a special set of linear
first order operators that involve reflections. From powers of these
Dunkl operators one can construct sufficiently many commuting operators
to render the problem integrable, in fact even superintegrable
(in rational/hyperbolic cases). Along with the multiplication by coordinates
and Weyl-reflections, Dunkl operators generate a (trigonometric) degeneration
of the so-called {\em double affine Hecke algebra}. The latter involves an
additional deformation parameter $q$ that is sent to $q=1$ when dealing
with the (undeformed) CS model. In order to understand the origin of the 
parameter $q$ one needs to turn to the rational Ruijsenaars-Schneider model  which is related to our hyperbolic CS model 
by a bi-spectral duality \cite{Ruijsenaars:1988pv}. Within the dual theory, $q$ controls the 
deformation from the rational to the hyperbolic version. Many more details 
on these topics can be found e.g.\ in \cite{Kirillov1995,Cherednik2005}.

All this structure is an integral part of the modern theory of
special functions. In the context of the (trigonometric) P\"oschl-Teller
problem we briefly sketched the relation between classical Jacobi and
q-deformed Askey-Wilson polynomials. The latter possess well developed
multi-variable extensions which are known as q-deformed HO
or (Macdonald)-Koornwinder polynomials $K$. Just as the trigonometric
P\"oschl-Teller problem can be solved through a degenerate limit of
Askey-Wilson polynomials, eigenfunctions of the trigonometric
CS Hamiltonian may be obtained from Koornwinder
polynomials in the limit $q\rightarrow 1$. This web of interrelations may be lifted from polynomials
to functions, i.e.\ from the trigonometric to the hyperbolic theory. The
lift turns Koornwinder or q-deformed HO polynomials into
what Rains refers to as {\em virtual} Koornwinder polynomials $\hat K$,
see \cite{Rains:2005}. We can also think of them as q-deformed
HO hypergeometric functions, up to normalization issues.

Before we can spell out a concrete formula we need to split the
data $\D,l$ that characterize the internal field into a partition
$(\l_1,\l_2)$ and a real parameter $\chi$. Upon imposing usual 
unitarity bounds, this is done as follows
\beqa
\l_2 & := & \lfloor \frac12(\D-l)\rfloor\\[1mm]
\l_1 := \l_2+l \quad & ,&  \chi  :=  \frac12(\D-l) - \l_2 \ .
\eeqa
The virtual Koornwinder polynomials $\hat K^{(2)}_{\l_1,\l_2}$ for
the root system {\it BC}$_2$ are functions of two variables $u_i, i=1,2$
that are associated to two-row partitions $(\l_1,\l_2)$. By appropriate choice of
their seven parameters, we can obtain conformal blocks as
\beqa \label{main}
& & (-4)^{-\D} \left(\frac{z\bar z}{16}\right)^a
G_{\D,l}(z_i)
= (u_1u_2)^{\chi+a}\,  \times \\[1mm]
& & \hspace*{-4mm}\lim_{q\rightarrow1^-}
\hat K^{(2)}_{\lambda_1,\lambda_2}
\left(u_i;q,q^{\epsilon/2},q^{-\chi-a};q^{a-b+1},-q^{a+b+1},1,-1\right)
\nonumber\eeqa
where  $\epsilon = d-2$ and
\beq
u_i  = -\frac{z_i}{(1+\sqrt{1-z_i})^2}
\eeq
for $i=1,2$ for $z_1 = z, z_2 = \bar z$ are obtained by inverting the
relations \eqref{zixirel} with $u_i = \exp{x_i}$. Note that the arguments
$u_i$ agree with the radial coordinates of \cite{Hogervorst:2013sma} up to
a sign. 

Virtual Koornwinder polynomials possess a binomial expansion in terms
of Okounkov's {\it BC}$_n$-type interpolation Macdonald polynomials,
see \cite{Rains:2005} section 7. These should be considered as
generalizations of the usual series expansion of hypergeometric
functions $F(u)$ in terms of monomials $u^k$. In the case the {\it BC}$_2$
root system, the interpolation Macdonald reduce to Gegenbauer polynomials
upon taking $q \rightarrow 1$ and the combinatorial prefactors may be
expressed through the hypergeometric functions $\ _4F_3$. The resulting
expansion reproduces a formula for conformal blocks
that was found by Dolan and Osborn in \cite{Dolan:2003hv}.

In order to demonstrate the powerful consequences of the relation between
conformal field theory and CS models we want to sketch a
few features of the blocks that seemingly were not observed before.
The first one concerns an interesting strong-weak coupling duality of the
{\it BC}$_n$ CS model that was noted by Serban in
\cite{Serban:1997cv} and can be neatly derived once the theory has been
q-deformed, using so-called Cauchy identities, see \cite{Rains:2005}.
The duality relates wave functions of the model with parameters
$(a,b,\epsilon)$ and
$$
(\tilde a,\tilde b,\tilde \epsilon) =
  \left(\frac{2(a+1)}{\epsilon}-1,\frac{2b}{\epsilon},
  \frac{4}{\epsilon}\right)
$$
This duality is non-perturbative in the integrable coupling $\epsilon$
that describes the deformation away from two decoupled P\"oschl-Teller
systems. Note that $\epsilon = 2$ is the self-dual point at which the
duality acts trivially. In the context of conformal blocks, the
parameter $\epsilon$ is related to the dimension $d$ through $\epsilon
= d-2$ and the self-dual point corresponds to 4-dimensional conformal
field theories.  Away from this special point, the duality allows to
write the blocks of a theory in dimension $d$ as an integral over
blocks of another theory in dimension $d'=2d/(d-2)$. An explicit 
formula for the integral kernel of this transformation may be 
inferred from \cite{Serban:1997cv,Rains:2005}. 

As we mentioned before, the q-deformation that proved useful at least
for the duality we discussed in the previous paragraph, originates from
the duality between the hyperbolic CS and the rational
RS model. Within the context of the RS model, one can also obtain
Gauss-like recurrence relations that describe the behavior of HO
hypergeometric functions under finite shifts of their parameters,
see e.g.\ \cite{vanDiejen}. There also exist growth estimates, see
\cite{Rosler} for some recent work, and an interesting relation with
(quantum) affine Knizhnik-Zamolodchikov equations for (q-deformed)
blocks \cite{matsuo1993,cherednik1992}. We will detail all
these features of blocks in a forthcoming longer paper.

\section{Conclusion and outlook}

The main observation of this work that the Casimir equation for
conformal blocks is equivalent to the Schr\"odinger equation
for the {\it BC}$_2$ CS model, embeds the central
objects in the bootstrap programme of $d$-dimensional conformal field
theory into the rich world of superintegrable quantum systems. The
deep connections to the modern theory of special functions have
powerful implications for conformal blocks of which we have
seen just two examples in the previous section.

There are a number of obvious extensions of our work that merit
further investigation. The first one concerns the extension to
superconformal field theory. In fact, the Casimir equations for
conformal blocks in superconformal field theories have
been discussed previously, but in most cases explicit formulas
are only known for a restricted set of external scalar fields,
see e.g.\ \cite{Bobev:2015jxa} for some interesting recent
developments and many references to the earlier literature.

Another interesting direction concerns the so-called crossing
kernel of $d$-dimensional conformal field theory. In the numerical
bootstrap program, the crossing symmetry is usually written in terms
of conformal blocks, with one side of the equation involving
blocks in the so-called $s$-channel while the other side
is expressed in terms of $t$-channel waves. The blocks in
the two different channels are related by the crossing kernel, so
that crossing symmetry may be expressed in terms of operator
product coefficients $\lambda_l(\D)$ and the crossing kernel,
stripping off the $(z,\bar z)$-dependent conformal blocks.
With the analytic control of conformal blocks we have
described above it is possible to obtain new and more explicit
formulas for the crossing kernels.

Let us finally stress again that in our entire discussion, the
dimension $d$ enters as a continuous parameter which is interpreted
as a coupling constant of the CS model. There
exist many conformal field theories for $d=2$ dimensions
that can be solved through their higher spin symmetries.
It should be possible to combine the results we outlined
above with the ideas that were put forward recently in
\cite{Alday:2015ota} to study the spectrum of conformal field
theories in $2+ \epsilon$ dimensions, at least for small
$\epsilon$. We will return to these interesting problems
in future work.
\medskip 

\noindent
{\bf Acknowledgements:} We wish to thank Ofer Aharony, Andrei Babichenko, 
Micha Berkooz, Martina Cornagliotto, Zohar Komargodski,
Madalena Lemos, Evgeny Sobko and in particular Gerhard Mack for interesting
discussions. This work was supported in part by the People
Programme (Marie Curie Actions) of the European Union's Seventh
Framework Programme FP7/2007-2013/ under REA Grant Agreement
No 317089 (GATIS), by an Israel
Science Foundation center for excellence grant, by the I-CORE program of the Planning
and Budgeting Committee and the Israel Science Foundation (grant number 1937/12), by
the Minerva foundation with funding from the Federal German Ministry for Education and
Research, by a Henri Gutwirth award from the Henri Gutwirth Fund for the Promotion of
Research, by the ISF within the ISF-UGC joint research program framework (grant
no. 1200/14) and by the ERC STG
grant 335182.

\bibliography{literatureCPW}

\end{document}